\documentstyle[preprint,aps]{revtex}             
\begin{document}
\preprint{UCLA-NT-9603} 
\draft

\title{\bf  Channel-Coupling Effects in High-Energy Hadron Collisions} 

\author { Chun Wa Wong}

\address{
Department of Physics and Astronomy, University of California, 
Los Angeles, CA 90095-1547}

\date{\today}

\maketitle

\begin{abstract}
The Two-Gluon Model of the Pomeron predicts strongly size-dependent 
high-energy hadron cross sections. Yet experimental cross sections for 
radially excited mesons appear surprisingly close in value. 
The strong coupling of these mesons in hadron collisions also 
predicted by the model permits a qualitative understanding of this puzzling 
behavior in terms of eigenmode propagation with a common eigen-$\sigma$. 
A detailed semiempirical coupled-channel model of the Pomeron is 
constructed to elucidate this and other features of high-energy hadron 
cross sections.
\end{abstract}


\narrowtext

  The Two-Gluon Model of the Pomeron (TGMP) has given a simple and 
rather successful picture of high-energy hadron-hadron scatterings  
\cite{Low75,Nus75}. Its most appealing feature is perhaps this: While 
accounting for the dependence on quark numbers emphasized so graphically 
by the additive quark model \cite{Lev65,Lip75,Don92}, it goes beyond 
the latter by giving a natural explanation of the flavor dependence of 
hadron cross sections as a size effect arising from the 
color separation inside colorless hadrons \cite{Gun77,Lev81}. This 
model prediction seems consistent with meson-nucleon ($mN$) cross 
sections for ground-state mesons \cite{Pov87}. Theoretical 
support has come from nonperturbative QCD \cite{Lan87} and 
lattice calculations \cite{Hen95}.

  The TGMP also predicts much larger $mN$ cross sections for 
radially excited mesons of larger radii. In the small-meson 
limit \cite{Gun77}, these cross sections are proportional to 
the meson ms radii, so that $\sigma_{\psi'N}$  
would be about four times larger than $\sigma_{(J/\psi)N}$.

  Yet their experimental values appear to be similar:  The older 
data on nuclear photoproduction, when analyzed in the 
Vector-Meson Dominance (VMD) model \cite{Bau78} 
gives a cross-section ratio of 0.77 (uncertainty 14) for the SLAC data 
\cite{Cam75}, 0.77 (9) for the NA14 data \cite{Bar87}, and 0.78 (11) for 
the E401 data \cite{Bin83}. [When corrected with modern branching ratios
\cite{RPP94}, these numbers should be changed to 0.79(15) and 0.86(10) 
for the first two data.] 
For muon production, one finds the ratio 0.81 (19) for the EMC data 
\cite{Aub83}, updated to 0.76 (18) with modern BR's. For the E772 data 
on the proton production in nuclei \cite{Ald91}, 
analyzed in the Glauber multiple-scattering formalism \cite{Won94} 
with uniform nucleon densities fitting experimental 
nuclear radii \cite{Dev87}, we get 9.0 mb/7.8 mb = 1.15. A more 
careful analysis of this rough equality, already noted in \cite{Won96},
has recently been made in \cite{Won96c}.

  Yet another piece of the puzzle comes from the nuclear production 
amplitudes of excited mesons relative to that of 
their ground state in $mN$ scatterings. H\"ufner and Kopeliovich 
(HK) \cite{Huf96} have recently deduced an experimental value of 0.46 
for the production amplitude ratio $\psi'/(J/\psi)$ in $p$-nucleus $(A)$ 
collisions \cite{Bag95,Ant93}. They have tried to 
explain this ratio in terms of a two-channel model of meson production 
using harmonic-oscillator (HO) wave functions and $mN$ cross 
sections in the small-meson limit. We shall show below 
that this 2-channel model is unstable, and that 
the generalization to many channels gives very poor results.

  The purpose of this paper is to show that this channel-coupling (CC) 
idea, when properly applied, seems to provide a key to a qualitative
understanding of these puzzling features. The essential 
ingredients seem to be: 
(1) using the full TGMP without making the small-meson approximation, 
(2) using more realistic meson wave functions, and 
(3) using many meson channels. A semiempirical CC model of the Pomeron 
is constructed to elucidate the many interesting features 
of $mN$ cross sections.

  Let us begin by pointing out that the coupling between different 
reaction channels of different hadron ($h$) excitations is unavoidable 
in $hA$ scattering. What is not clear is whether the coupling 
is strong or weak in the TGMP. We shall see that it is strong for 
color-singlet mesons on nucleons, 
but weak for color-octet mesons.

  For definiteness, we consider the simplified situation where 
the target hadrons are the nucleons in a nucleus and where only the 
projectile meson undergoes radial excitations. Channel coupling is 
in general a multiple-scattering phenomenon. Its physics becomes 
particularly transparent when described in the attenuation 
appoximation \cite{Ger88,Huf96}, where 
a multi-channel wave is attenuated by the matrix 
exp$(-\rho_0 L \sigma)$ in channel space. Here $\rho_0$ is the nucleon 
density, $L  =  \mbox{$3\over 4$}R(1-1/A)$ is the average attenuation 
length  across a nucleus of radius R, and $\sigma$ is the matrix 
containing both single- and cross-channel absorption cross sections. 
It is usefully expressed as the dimensionless matrix 
$M = \sigma/\sigma_0$, where $\sigma_0 = \sigma_{11}$, 
so that $M_{11} = 1$.

   The attenuation of the multi-channel hadron wave can then be 
understood in terms of the 
eigenvalues and eigenvectors of $M$. After a generally 
complicated transient, the wave will eventually decay into that 
eigenmode belonging to the smallest eigen-$\sigma$, hereafter called 
the ``lowest'' eigenmode. When produced in an eigenmode, the hadrons
will propagate in that eigenmode over all distances,
with the same eigen-$\sigma$, and therefore the same attenuation, 
in all channels. However, only the lowest eigenmode is stable 
against perturbations.

   In the HK model \cite{Huf96}, the nucleons remain unexcited, 
but the propagating $J/\psi$ meson can be excited radially. In 
addition, the CC matrix $M$ is calculated in the 
small-meson limit using HO wave functions. It is then a
tridiagonal matrix in channel space. 

   The HK model is based on the observation that in the 2-channel 
approximation, the theoretical $\psi'/\psi$ ratio $F$ of wave 
amplitudes in the lowest eigenmode agrees with the experimental 
value of 0.46 (6). This agreement is displayed in the first row of 
Table I, where $\Sigma_i$ is the $i$th eigen-$\sigma$, and ${\bf v}_i(j)$ 
is the wave amplitude in the $j$th channel of its eigenvector.

  However, as the number $n$ of channels increases beyond 2, the
amplitude ratio $F$ increases rapidly to above 1, thus destroying 
the agreement with experiment. At the same time, $\Sigma_1$ decreases 
towards zero, making the nucleus increasingly transparent to meson 
propagation. This result also contradicts the well-known experimental 
fact that the effective $\psi N$ cross section in nuclei is 
larger than its value in free space \cite{Ger88}.

  The table shows that in this HO $r^2$ model, stability in the 
$\psi'/\psi$ amplitude ratio $F$ requires the inclusion of at least 
30 channels. The results numerically extrapolated to $n=\infty$ by 
using a ``diagonal'' rational approximant \cite{Pre89} are also given.

   The situation is more interesting if the small-meson 
approximation is not used. We start with the full $hN$
amplitude of the TGMP first derived by Low and others 
\cite{Low75,Nus75,Gun77,Lev81,Lan87,Dol92,Duc93}, using either 
perturbative (P) or nonperturbative (NP or Cornwall) gluon propagators
\cite{Cor82,Hal93} with an effective gluon mass $m=m(q^2=0)$. 
Application of the optical theorem yields the total $hN$ 
cross sections

\begin{eqnarray}
\sigma_{ij}(h)=8 n_h n_N \alpha_s^2 
\int d^2{\bf k} D^2(k) \Phi_{ij}(h;k) \Phi_N(k)\,,
\end{eqnarray}
where $n_i$ is the number of quarks in hadron $i$, and

\begin{eqnarray}
\Phi_N(k) \simeq 1-f_N(3 k^2)\,,
\end{eqnarray}

\begin{eqnarray}
\Phi_{ij}([q \bar q]_1;k)=\delta_{ij}-f_{ij}(4 k^2)\,,
\end{eqnarray}

\begin{eqnarray}
\Phi_{ij}([q \bar q]_8;k)=\delta_{ij}+\mbox{$1\over 8$}f_{ij}(4 k^2)\,,
\end{eqnarray}

\begin{eqnarray}
\Phi_{ij}([(q \bar q)_8g]_1;k) \simeq \mbox{$9\over 4$}[\delta_{ij}
-f_{ij}(4 k^2)]\,.
\end{eqnarray}
Here $f_{ij}$ is the diagonal or off-diagonal wave-function form 
factors. The proton and $[q \bar q]_8$ expressions are from 
\cite{Dol92}, while that color factor for the hybrid is from
\cite{Liu94,Kha95}. The expression for hybrid mesons involves only 
the separation between g and $(q \bar q)_8$ if the weak dependence 
on the $q \bar q$ separation is neglected.

  Each hadron factor in Eqs. (2-5) contains two terms: 
(1) a one-body term diagonal in the channel index caused by 
the exchange of both Pomeronic gluons with the same body: $q$, $\bar q$, 
or the $(q\bar q)_8$ of the hybrid treated as a single body, and 
(2) a two-body term arising from the gluons interacting with both 
parts of the hadron, dependent on form factors, and responsible
for channel coupling. For a 
colorless hadron, these terms interfere destructively because the 
scattering amplitude vanishes for a point hadron \cite{Gun77}. 
This is why the cross sections are so sensitive to hadron sizes, and 
could be quite small even though each term is large. In color octets 
however, the two-body term is much weaker and adds 
constructively to the one-body term. The result is much larger cross 
sections and much weaker channel couplings.

  Ratios of these cross sections are theoretically simpler: They are 
independent of $\alpha_s$ in the P treatment, and only weakly dependent 
on the QCD energy scale $\Lambda_{QCD}$ in the NP treatment. 
We determine $m$ by a best fit to the ratio $\pi N/pN$ and $KN/pN$ 
of the experimental Pomeron-exchange cross sections at the experimental 
hadron matter radii, both given in Table II. (Matter radii are deduced 
from charge radii.)
The result is $m = 0.08(10)$ GeV in the P treatment, and 0.27(6) GeV 
in the NP treatment used below, the latter being at the lower end of the
Cornwall estimate of 0.5(3) GeV \cite{Cor82}. These masses are obtained 
with dipole form factors and $\Lambda_{QCD}=0.3$ GeV, but the results 
for Gaussian form factors and other values of $\Lambda_{QCD}$ are 
practically the same. Note that the product $\alpha_sD(k)$ does not 
depend on $\alpha_s$ in the NP treatment, and that the estimated error 
in $m$ comes only from the errors in the input hadron radii, since the 
errors in the Pomeron cross sections are unknown.

  Meson wave functions are needed to calculate the CC matrix $M$. 
The effort is greatly reduced by using HO and the hydrogenic 
(Hy) wave functions. Neither turn out to be adequate, but they bracket 
more realistic wave functions, thereby allowing a simulation of the 
latter by the interpolation of their $M$ matrices. As in Ref.\ \cite{Huf96},
we have used $\sigma_{total}$ instead of $\sigma_{abs}$ in $M$. 

   As shown in Table I for $J/\psi$ mesons, the results for HO wave 
functions, though improved over the HO $r^2$ model, are not yet 
convergent with four channels. On the other hand, the 
results for Hy wave functions are beginning to converge with four 
channels, because the off-diagonal matrix elements are much 
weaker. This difference is also responsible for the result, 
already noticeable in Table I, that channel mixing in the lowest 
eigenmode becomes much weaker in the Hy model. (The channel components  
in the lowest eigenmode are highly coherent and have the same sign 
because all the off-diagonal matrix elements of $M$ have the same negative 
sign.) 

  The wave-amplitude ratio $F$ for the first two channels turns out to 
be too large for the HO model, but too small for the Hy model, when 
compared to the experimental value of 0.46(6) quoted previously.  
To simulate more realistic wave functions, we take that linear 
combination of the two theoretical cross-section matrices which will 
reproduce the experimental value of $F$ when the $n=2,3,4$ results 
are numerically extrapolate to $n=\infty$ \cite{Pre89}. This gives the  
``75\%(HO)/25\%(Hy)'' model shown in Table III. Obviously 
the fit would change with future changes in $F$, but the qualitative 
features of the model should be similar.

   The lowest eigen-$\sigma$ $\Sigma_1$, which controls the  
steady-state propagation, should be lower than the smallest 
single-channel cross section $\sigma_0$. It is about $0.64\sigma_0$ 
for $J/\psi$ mesons in the 75/25 model. As channel-coupling 
decreases in going from the HO $r^2$ to the Hy model, the reduction of 
$\Sigma_1$ below $\sigma_0$ also becomes less and less. 

  The Pomeron-exchange parts $\sigma_0$ of ground-state 
$hN$ cross sections are themselves of interest. 
These can be obtained from the theoretical cross-section ratios by 
multiplication into the empirical $pN$ Pomeron contribution of 
$21.70 (s/{\rm GeV}^2)^{0.0808}$ mb fitted by Ref.\ \cite{Don92}. In this 
way, we not only reduce the considerable sensitivity of each cross section 
to model parameters, but also recover the experimental $s$ dependence. 
Using dipole form factors fitting theoretical meson ms radii, we obtain 
the TGMP predictions shown in Table III. The errors shown come from the
uncertainty of the gluon mass $m$. In addition, the cross sections are 
smaller by about 5\% for Gaussian form factors.

  At $\sqrt{s}=20$ GeV, we find $\sigma_0 \simeq 5.2$ mb for a 
$J/\psi$ meson ms radius of 0.044 fm$^2$ \cite{Buc81}.  This is 
about twice the value of 2.4 mb calculated in the gluon-fusion 
model of \cite{Kha94}, and almost the same as the 6 mb obtained in 
\cite{Ger88}, or the 8 mb reported here, both from nuclear 
suppression data. If the meson ms radius is 0.055 fm$^2$ \cite{Eic80},
$\sigma_0$ would have been 6.2 mb. 

  Stable asymptotic propagation in nuclei on the other hand involves 
the lowest eigen-$\sigma$ $\Sigma_1$. For $J/\psi$ 
mesons, this is $\simeq 0.64 \sigma_0 =$ 3.3(8) mb. The additional 
uncertainties from 
$\Delta F$ or the extrapolation to infinite matrix dimension  
probably does not exceed 0.3 mb each.

    A number of conclusions can be drawn:
(1) Colorless mesons of different radial excitations and sizes 
experience the {\it same} total $mN$ cross section in nuclear 
suppression if they are propagating close to an eigenmode, usually 
the lowest eigenmode. 
(2) The smallest eigen-$\sigma$ $\Sigma_1$ is smaller 
than $\sigma_0 = \sigma_{11}$ of the meson ground state. 
(3) For colorless $J/\psi$ mesons, $\Sigma_1$ is smaller than the 
effective $mN$ cross section deduced from nuclear suppression. 
(4) The single-channel cross section $\sigma_0$ is about twice the 
2.4 mb given by the gluon-fusion model. Given the many 
uncertainties of our Pomeron model, it is not clear if the 
discrepancy is serious, and if so how it should be understood. 
Channel-coupling effects should also be present in the gluon-fusion 
model, but it is not known if they are strong or weak.
(5) The nuclear propagation of hybrid mesons is quite similar to that 
for colorless mesons, but their cross sections are larger by the 
relative color factor \mbox{$9\over 4$}. On the other hand, color 
octets of different sizes tend to propagate independently in nuclei 
with roughly the same cross section. This is because the 
size-dependent, channel-coupling 2-body term is only \mbox{-$1\over 8$} 
of that in color singlets.
 
   Concerning conclusion (3), it has been suggested that the the larger 
cross sections of the absorption model could arise from the appearance 
of $J/\psi$ mesons in color octet forms, either together with a 
gluon in a hybrid \cite{Kha95}, or as bare color octets \cite{Won96b}. 
A hybrid explanation of the increasing apparent nuclear absorption with 
increasing $x_F$ discussed in \cite{Won96b} can readily be given in our 
TGMP, with the ms hybrid radius increasing from about 0.03 fm$^2$  
at small $x_F$ to about  0.25 (16) fm$^2$ at $x_F \simeq 0.6$. 

  Conclusions (1) and (2) have interesting implications in the 
generalized VMD model of meson photoproduction 
from nuclei. In this model, the photon appears as a coherent admixture 
of all possible virtual vector mesons. Progress in this problem in the 
past has been hindered by the lack of information on both the 
channel-coupling matrix and the incident meson amplitudes \cite{Bau78}. 
Our CC model gives very specific predictions, although limited to only 
a small number of channels in the present calculation. The lowest 
eigenvectors of our largest $n=4$ models, all with 75/25 mixing of wave
functions, are shown in Table III for several meson families. The 
ground-state component ${\bf v}_1(1)$ has been taken to be 1 for ease 
of comparison. Each eigenvector can be compared with the GVMD input 
amplitude vector ${\bf v}_\gamma = (1,f_1/f_2, ... )$, where $f_i$
is the universal meson coupling constant to its source 
obtained from \cite{RPP94}, and for $\rho'$, from \cite{Cer73}. 

  We see that the input amplitude vector for photoproduced mesons is 
rather close to the lowest eigenvector, especially for $J/\psi$'s. 
Hence mesons are produced close to this eigenmode, and
propagate in nuclei with roughly 
the same reduced cross section $\Sigma_1$ in different radial 
excitations. However, this reduction is not enough to explain why 
$\sigma_{(J/\psi)N}$ is only 1.8-1.9 mb at $\sqrt{s}=$ 18-20 GeV when 
deduced from photoproduction data under traditional VMD \cite{Won96c} 
unless a smaller gluon mass is used. 

The experimental ratio $\sigma(\psi')/\sigma(J/\psi)$ for production 
in $pA$ collisions is also independent of $A$ \cite{Bag95}, meaning 
that the hadron production at a nucleon is also close to the lowest 
eigenmode when interpreted in the coupled-channel model. 
This idea of ``eigenmode production", assumed both here and in 
\cite{Huf96}, will required detailed justification.  

I would like to thank Dr. Cheuk-Yin Wong for many helpful discussions.

\begin{table}
\caption{The lowest eigenmode of the channel-coupling matrix $M$ for 
$J/\psi$ mesons in different models.}
\begin{tabular}{rrlrlrl}
Model&  HO $r^2$&  &  NP/HO&  &  NP/Hy&    \\  
$n$&  $\Sigma_1/\sigma_0$&  $F$*&  $\Sigma_1/\sigma_0$&  $F$*&
      $\Sigma_1/\sigma_0$&  $F$*  \\
\hline
2&         0.61&  0.47&  0.66&  0.53&  0.92&  0.142  \\
3&         0.44&  0.68&  0.49&  0.72&  0.89&  0.153  \\
4&         0.35&  0.80&  0.40&  0.83&  0.88&  0.158  \\
15&        0.10&  1.10&  &  &  &    \\
30&        0.05&  1.16&  &  &  &    \\
$\infty$&  0.00&  1.22&  0.02&  1.22&  0.86&  0.169
\end{tabular}
*$F={\bf v}_1(2)/{\bf v}_1(1)$
\label{table1}
\end{table}

\narrowtext

\begin{table}
\caption{Pomeron-exchange contribution (in mb) to 
$\sigma_\protect{hN}=X_\protect{hN} s^\protect{0.0808}$ at 
$\protect\sqrt{s}=20$ GeV using dipole form factors.
\label{table2}}
\begin{tabular}{lllddd}
Hadron&          $\langle r^2 \rangle $*&  Ref&  $X_{hN}$&       &
$\sigma_{hN}$  \\ 
$D(k)$&                  &     &        NP&         P&   NP            \\  
\hline
$p$&              0.67(2)&  \cite{Bor75}&  21.70$\dagger$ &  &
35.21$\dagger$  \\
$\pi$&            0.44(1)&  \cite{Ame86}&  13.63$\dagger$ &  &
22.12$\dagger$  \\
$K$&              0.31(5)&  \cite{Ame86a}& 11.82$\dagger$ &  &
19.18$\dagger$  \\  
\hline
$J/\psi$&           0.044&  \cite{Buc81}&  3.2(7)&    3.7(5)&  5.2(12)  \\
&                   0.055&  \cite{Eic80}&  3.8(8)&    4.3(6)&  6.2(13)  \\

$\psi'$&            0.181&  \cite{Buc81}&  8.6(9)&    8.8(7)&  14.0(15)  \\
$\Upsilon$&         0.013&  \cite{Buc81}&  1.1(3)&    1.6(3)&  1.9(6)  \\
$\Upsilon'$&        0.063&  \cite{Buc81}&  4.2(9)&    4.7(6)&  6.8(14)  \\
$\rho, \omega$&      0.54&  \cite{God85}& 14.90(12)& 14.91(9)&  24.2(2)  \\
$K$*&                0.37&  \cite{God85}& 12.6(4)&    12.6(3)&  20.5(7)  \\
\hline
$\rho_8$&            0.54&  \cite{God85}&   28(8)&          &  45(12)  \\
$(J/\psi)_8$&       0.044&  \cite{Buc81}&   29(8)&          &  48(13)  \\
$\Upsilon_8$&       0.013&  \cite{Buc81}&   29(8)&          &  47(13)
\end{tabular}
 *  In fm$^2$.   $\dagger$  Experimental results from \cite{Don92}.
\end{table}

\narrowtext

\begin{table}
\caption{The lowest eigenmode in the 75\%(HO)/\-25\%\-(Hy) model and 
the GVMD input vector ${\bf v}_\gamma$ in different meson families.
Meson ms radii used are those of Table II.}
\begin{tabular}{llll}
Mesons&
Type&  $\Sigma_1/\sigma_0$&  ${\bf v}_1$ or ${\bf v}_\gamma$  \\
\hline
$\rho$&        $n=4$&  0.69&  $(1, 0.69, 0.50, 0.35)$  \\
&         $n=\infty$&  0.46&  $(1, 0.90, ...)$  \\
&           $\gamma$&      &  $(1, 0.35, ...)$  \\
\hline
$J/\psi$&      $n=4$&  0.71&  $(1, 0.40, 0.19, 0.09)$  \\
&         $n=\infty$&  0.65&  $(1, 0.45, ...)$  \\
&           $\gamma$&      &  $(1, 0.58,0.32, 0.25, ...)$  \\
\hline  
$\Upsilon$&    $n=4$&  0.74&  $(1, 0.32, 0.12, 0.05)$  \\
&         $n=\infty$&  0.70&  $(1, 0.35, ...)$  \\
&           $\gamma$&      &  $(1, 0.64, 0.58, 0.40, 0.45, 0.29,...)$
\end{tabular}
\label{table3}
\end{table}

\end{document}